\documentstyle[aps,prl]{revtex}
\begin{document}
\draft
\preprint{UM-P-96/33, CfPA-96-TH-10}
\title{The mixmaster universe is chaotic}
\author{ Neil J. Cornish$^{\star}$ and Janna J. Levin$^{\star\star}$} 
\address{$^{\star}$School of Physics, University of Melbourne, Parkville 3052,
Victoria, Australia}
\address{$^{\star\star}$Center for Particle Astrophysics,
UC Berkeley, 301 Le Conte Hall, Berkeley, CA 94720-7304}
\twocolumn[
\maketitle
\widetext
\begin{abstract}
For the past decade there has been a considerable debate about the
existence of chaos in the mixmaster cosmological model.
The debate has been hampered by the coordinate, or observer dependence
of standard chaotic indicators such as Lyapunov exponents. Here we
use coordinate independent, fractal methods to show the mixmaster
universe is indeed chaotic.
\end{abstract}
\pacs{05.45.+b, 95.10.E, 98.80.Cq, 98.80.Hw}
]
\narrowtext

The origin of the universe and the fate of collapsing stars 
are two of the great mysteries in nature.
In general relativity without exotic matter, 
the singularity theorems of Hawking and Penrose \cite{hp}
argue that the gravitational
collapse of very massive stars ends singular and that
the universe was born singular.
These singular settings force gravity to face quantum mechanics.
As well as exposing the fundamental laws of physics,
the singular cores of black holes and the origin of the cosmos
draw deep connections to the
laws of thermodynamics and, as we will discuss here,
to chaos.

Earlier, Khalatnikov and Lifshitz\cite{kl1} 
argued that singular solutions were the exception rather than the rule,
putting them at odds with the singularity theorems. 
Their argument was that deformations in spacetime would
be amplified during collapse and consequently would fight the formation
of a singularity.
This implied that the known symmetric singular solutions were
atypical. The conflict was resolved when
they realized the singularity in a collapsed star could be chaotic
\cite{{kl2}}. They conjectured that
a generic singularity drives spacetime to churn and oscillate chaotically.
Independently, Misner \cite{mix} suggested a chaotic approach
to an early universe singularity.  In his mixmaster universe, the
different directions in 3-space alternate in cycles of anisotropic collapse
and expansion. A popular account of these developments may be
found in Thorne's recent book\cite{kip}.

While the emergence of chaos helped our understanding 
of singularities in general relativity,
building a resilient theory of relativistic chaos 
has become a task of its own. A debate has raged over whether or not
the mixmaster universe is chaotic. Studies of the mixmaster
dynamics using both approximate maps\cite{barrow2,bt} and numerical
integrations\cite{fm,pullin} have each yielded conflicting results
as to the existence of positive Lyapunov exponents -- a
standard chaotic indicator.
Finally, it was realised\cite{pullin,rugh} that
Lyapunov exponents are coordinate dependent and the conflicting results
were a consequence of the different
coordinate systems. In short, Lyapunov exponents are not reliable
indicators of chaos in general relativity. 
Using a different approach, it was shown that
the mixmaster equations fail the Painlev\'{e} test\cite{pan}.
This suggests that the mixmaster may be chaotic, but the Painlev\'{e}
test is also inconclusive.
A detailed review of the mixmaster
debate can be found in Ref.\cite{burd}. 

In this letter we show that the mixmaster universe is indeed chaotic
by using {\em coordinate independent}, fractal methods. 
A fractal set of self-similar universes is uncovered by numerically
solving Einstein's equations. These universes form 
fractal boundaries in the space of initial conditions. Such fractal
partitions are the result of chaotic dynamics.
We emphasize that our approach can be used to study
any system described by general relativity.
The mixmaster is studied here as a topical example.

The mixmaster universe\cite{mix}
has closed spatial sections with the topology of a three sphere. The vacuum
field equations of general relativity lead to the equations 
\begin{equation}\label{mixeq}
(\ln a^2)''=(b^2-c^2)^2-a^4 \quad {\it et\; cyc.}\; (a,b,c) \; .
\end{equation}
Here $(a,b,c)$ are the scale factors for the three spatial axes,
a prime denotes $d/d\tau$ and $dt=abc\, d\tau$. In a numerical
study it is advantageous to use $\alpha=\ln a$, $\beta=\ln b$, $\gamma=
\ln c$ and $dT=d t /(abc\ln(abc))$ as integration variables. These variables
cautiously approach the curvature singularity at $t=0$. 
Asymptotically the new time variable is related to the comoving time $t$ by
$T=\ln(\ln (1/t))$. 
The equations of motion (\ref{mixeq}) can be integrated 
to yield the Hamiltonian constraint $H=0$, where
\begin{eqnarray}\label{ham}
&&H=(\ln a )'(\ln b)'+(\ln a)'(\ln c)'+(\ln b)'(\ln c)'\nonumber \\
&&\quad -\frac{1}{4}\left[a^4+b^4+c^4
-2\left(a^2 b^2 + b^2 c^2+ a^2 c^2\right)\right] \, .
\end{eqnarray}
When the potential terms on the r.h.s. of (\ref{mixeq}) are small, the
mixmaster coasts in an approximate Kasner phase described by
\begin{equation}\label{line}
ds^2=-dt^2+t^{\,2{\textstyle p}_{{\scriptstyle a}}}dx^2
+t^{\,2{\textstyle p}_{{\scriptstyle b}}}dy^2+
t^{\,2{\textstyle p}_{{\scriptstyle c}}}dz^2 \, .
\end{equation}
The indices can be written as
$p_1=(1+u)/(1+u+u^2)$, $p_2=-u/(1+u+u^2)$, and $p_3=(u+u^2)/(1+u+u^2)$,
where $(a,b,c)$ can take any ordering of $(1,2,3)$ and $u\in[1,\infty)$.

Initial
conditions can be set on the surface 
$\alpha=0$,
$d\alpha/d T=\dot{\alpha} <0$, with
the four variables
$(u,v,s,\Sigma)$\cite{barrow1}:
\begin{eqnarray}\label{kas}
&&\alpha=0\, , \hspace{0.2in} \beta={\Sigma \over 1+v+uv} ,\hspace{0.1in}
\gamma ={ \Sigma (v+uv)\over 1+v+uv}, \nonumber \\
&&\dot{\alpha}=s \Sigma p_{1}\, , \hspace{0.2in} \dot{\beta}=s\Sigma p_2 \, ,
\hspace{0.2in} \dot{\gamma}= s\Sigma p_3\, .
\end{eqnarray}
The anisotropy in the sizes and 
velocities of the axes is quantified by $v$ and $u$.
The variables $\Sigma$ and $s$ are overall scale factors.
We will
primarily be interested in the evolution of $(u,v)$ as the mixmaster
singularity is approached. That is, we are mostly interested in
the relative rates of expansion of the three spatial directions.

Before turning to the full mixmaster dynamics, we consider
the properties of a two dimensional map that approximately describes
the evolution of $(u,v)$ as the singularity is approached. 
We use the map to guide our study of the full dynamics.
A detailed comparison of our treatment and the original
studies of the related Gauss map~\cite{barrow2,kl2}
will be delineated elsewhere~\cite{us}.
We mention that unlike previous treatments, we focus on 
chaotic scattering.
A complete description of
chaotic scattering is encoded in the unstable periodic
orbits~\cite{bl,ott}. These orbits form what is know as a
{\em strange repellor} -- a close cousin of the more familiar
strange attractor.
We expose the fractal nature of the repellor.  
The same fractal set is then found
in the full dynamics.

The map evolves $(u,v)$ forward in discrete jumps\cite{kl2,barrow1}, 
\begin{equation}
	F(u ,v )= \left \{\begin{array}{lll}
	u-1 ,&\ v +1\, ,  &   u \geq 2 \quad ({\rm oscillations}) \\ & & \\
{\displaystyle {1\over u -1}},&{\displaystyle {1\over v} +1 }\, ,&   
u< 2 \quad {\rm (bounces)} \, .
	\end{array} \right.
	\label{map}
\end{equation}
During an oscillation, one pair of axes oscillates out of
phase while the third decreases monotonically. At a bounce, the roles of the
three axes are interchanged and a different axis decreases monotonically.

The strange repellor is the fractal set of points along periodic orbits, 
$F^k(\bar u,\bar v)=(\bar u,\bar v)$. 
Physically these orbits are self-similar universes.
Periodic orbits 
with period $p\leq k$ can be divided into $m$ oscillations and $k-m$ bounces.
Since bounces and oscillations do not commute,
the number of $u$ fixed points on an orbit with 
$m$ oscillations is $^kC_{m}$.
The total number of fixed points 
at order $k$ is given by
the sum over all possibilities squared,
$N(k)= (2^k-1)^2$.
Thus, the topological
entropy\cite{ott} of the strange repellor is given by
\begin{equation}
H_{T}=\lim_{k\rightarrow \infty} {1 \over k} \ln N(k) = 2\ln 2    \, .
\end{equation}
Since $H_{T}>0$ the $F$-map is chaotic.

To make contact with the continuum dynamics, we display a portion of
the repellor's future invariant set in Fig. 1. The future invariant set
is the collection of lines $F^k(\bar u)=\bar u$, ($v$ arbitary). 
The sequence of gaps around the rationals form what is known as a 
Farey tree.  In a complementary fashion, the repellor lies on the
periodic irrationals of the irrational Farey tree \cite{us}.
A similar
collection of lines occurs in each integer interval $u=[n,n+1]$, but
with exponentially decreasing density.
The fall off 
can also be understood from the combinatorics of 
oscillations $O$ and bounces $B$. 
A root in this interval corresponds to the word,
$O^{(n-1)}B$. 
The number of $n$-letter words that can be formed from a 2-letter
alphabet is $2^{n}$, so the fraction of roots
in each interval is $2^{-n}$.
The map and
the continuum dynamics will be seen to describe the same future invariant set.

The universes which comprise the strange repellor forever
repeat some prescribed cycle in $(u,v)$. In contrast, a typical mixmaster
universe will launch into an infinite pattern of oscillations and bounces
that never repeats. Moreover, typical universes have an invariant
probability $\sim u/\ln u$ that grows with $u$\cite{berger}. As a result,
typical universes scatter to $u\rightarrow \infty$ while universes on
the repellor are concentrated at small $u$ values.

\
\begin{figure}[h]
\vspace{55mm}
\includegraphics{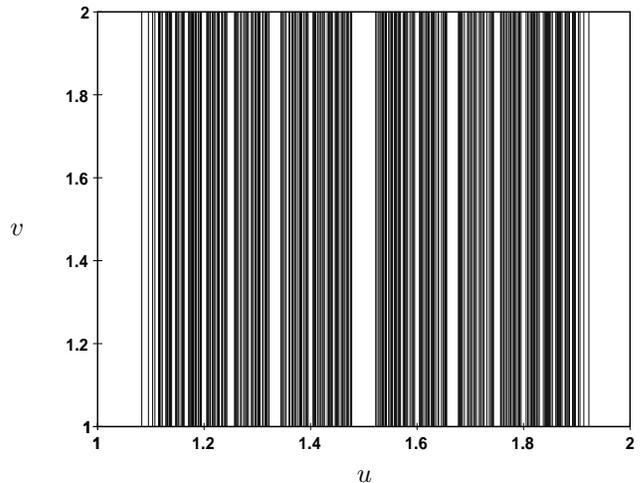}
\vspace{8mm}
\caption{ The map's future invariant set in the interval $1<u<2$.}
\end{figure}

\begin{picture}(0,0)
\put(-5,140){${v}$}
\put(126,47){$u$}
\end{picture}

\vspace*{-5mm}

We can quantify the multifractal nature of the
repellor by measuring its fractal dimensions $D_{q}$, where\cite{ott}
\begin{equation}\label{dimdef}
D_{q}={1 \over q-1} \lim_{\epsilon \rightarrow 0} {\ln 
\sum_{i=1}^{N(\epsilon)} (p_{i})^{q} \over \ln \epsilon } \; .
\end{equation}
Here $N(\epsilon)$ are the number of hypercubes of
side length $\epsilon$
needed to cover the fractal and $p_{i}$ is the fraction of points
in the $i^{{\rm th}}$ hypercube. 
The standard capacity dimension is
recovered when $q=0$, the information dimension when $q=1$, etc.
For homogeneous fractals all the various
dimensions yield the same result. The dimensions $D_{q}$ are
invariant under diffeomorphisms for all $q$. 

Since the periodic orbits of (\ref{map}) are everywhere
dense, it follows that the future invariant set in Fig. 1 has $D_{0}=2$.
However, points on a small period orbit are visited with greater
frequency and so have a larger $p_i$ than high period orbits.  
This generates an
uneven distribution which
ensures that $D_{q}<2$ for $q> 0$. Numerically solving for
all roots up to $k=16$ we find $D_{1}=1.87\pm 0.01$, thus confirming
the multifractal nature of the repellor.

If the $F$-map had been obtained from the full Einstein equations
without approximation, we could conclude that the
mixmaster universe is chaotic. Since approximations were
made\cite{kl2,barrow1}, the possibilty remains that
the full equations are integrable. The approximations may have failed
to preserve some integrals of the motion, thus leading to a false
chaotic signal. We show this is not the case.

Since the mixmaster phase space is not compact, any chaotic behaviour
is likely to be transient. There is a standard procedure to investigate
such chaotic scattering. First we identify the different asymptotic outcomes
the system might have. Once outcomes are identified,
several methods can be used to search for a strange
repellor. The simplest method looks for a fractal pattern in
plots of an appropriately defined scattering angle and impact
parameter. Alternatively, the strange repellor can be hunted directly with
a numerical shooting procedure called PIM triples\cite{pim}.

Our prefered method is to look for fractal basin boundaries\cite{greb}.
Each outcome has a basin of attraction in the space of initial conditions.
We may plot these basins by assigning a different color to each outcome,
and then coloring all initial conditions according to their outcome.
If the boundaries between these outcomes
are smooth, then the dynamics is regular. Conversely, if the boundaries are
fractal, the dynamics is chaotic. The set of points belonging to the
fractal boundary form the strange repellor's future
invariant set.

The power of these methods as a tool for studying chaos in general
relativity is twofold. First,
a fractal is a non-differentiable structure and so cannot be removed
by any differentiable coordinate transformation.
Thus,
a fractal basin boundary provides an observer independent signal
of chaos.
Second, most systems in general relativity
have non-compact phase spaces, so most chaos will be transient.
Other coordinate independent methods of studying
chaos in general relativity, such as curvature based methods\cite{curve},
only work for compact systems.  Previously fractal
techniques were used to show
there is chaos in multi-black hole spacetimes\cite{carl},
and in various inflationary cosmological models\cite{cos}.

As it stands, the mixmaster dynamics rarely leads to
definite outcomes. For typical trajectories the sequence of oscillations
and bounces continues {\it ad infinitum} as the singularity is approached.
A trajectory that visits any finite value of $u$, no matter how large,
will eventually return to bounce again.
The way around this asymptotic backwash problem is to
assign outcomes at some large, but finite distance from the scattering
region. This procedure cannot lead to a false chaotic signature. At worst,
it will return smooth boundaries for a chaotic system. This occurs when
trajectories near the strange repellor are prematurely assigned to a
particular outcome. 

For the mixmaster universe, typical trajectories scatter to large
$u$ values, while orbits on the repellor are concentrated at small
$u$ values. Accordingly, we assign outcomes when $u$ becomes larger
than some fixed number, $u_{{\rm max}}$, during a Kasner phase.
Due to the system's $SO(3)$ symetry, a large $u$ value leads to
three equally likely outcomes. Physically, these are the three states
of highly anisotropic expansion
with $p_1\approx p_2\approx 0$ and $p_3\approx 1$. Thus, the space of
initial conditions can be color coded depending on which
axis is collapsing most quickly. We color these black for $a$,
grey for $b$ and white for $c$.

\
\begin{figure}[h]
\vspace{56mm}
\includegraphics{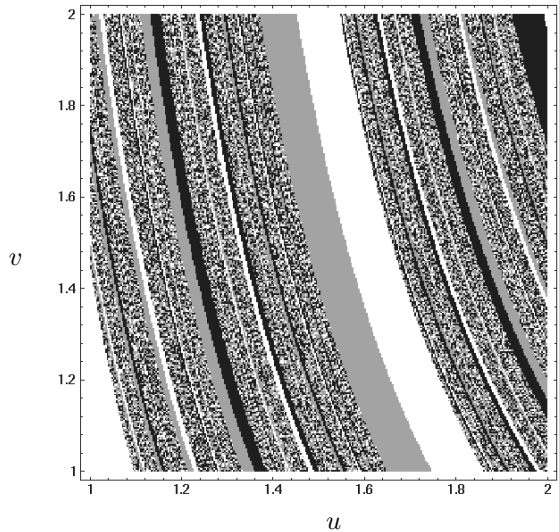}
\vspace{15mm}
\caption{Basin boundaries in the $(u,v)$ plane for the full,
unapproximated mixmaster dynamics.}
\end{figure} 

\begin{picture}(0,0)
\put(5,150){$v$}
\put(125,50){$u$}
\end{picture}

\vspace*{-4mm}

Our prescription is very easy to implement numerically. In order to
faciliate comparison with Fig.~1, we chose initial condition in accordance
with (\ref{kas}) by selecting $(u_o,v_o)$ from a $300^2$ grid,
setting $s_o=1$ and then using (\ref{ham}) to fix
$\Sigma_o$. The initial conditions are then evolved
according to the equations of motion (\ref{mixeq}), and the ratios of
$\dot{\alpha}$, $\dot{\beta}$ and $\dot{\gamma}$ are monitored to
see if the universe is in an approximate Kasner phase described
by (\ref{line}). If it is, the value of $u$ is extracted and an outcome is
assigned if $u>u_{{\rm max}}$. We choose $u_{{\max}}=7$ since there is only
a 1 in $2^6$ chance that a trajectory with $u>7$ lies on the strange repellor.
Moreover, typical aperiodic trajectories will scatter to $u>u_{{\rm max}}$
after a few bounces, so the numerical integration is kept short
and numerical errors do not become large. To confirm this,
the unenforced Hamiltonian constraint (\ref{ham}) was monitored at
all times and found to be satisfied within numerical tolerances.

In Fig. 2 we display the basin boundaries in a portion of
the $(u,v)$ plane. We see a complicated mixture of both
regular and fractal basin boundaries. The numerically generated
basin boundaries are built of universes which ride the repellor
for many orbits before being thrown off. Similar fractal basins can be
found by viewing alternative slices through phase space, such
as the $(\beta, \dot\beta)$ plane.
The overall morphology of the basins is altered little by
demanding more strongly anisotropic outcomes. From Fig. 3 we see that
the fractal nature of the boundary persists on finer and finer scales.

Aside from some mild warpage, the future invariant set (basin
boundaries) seen in Figs.~2~\&~3 for the full dynamics is
strikingly similar to that shown in Fig.~1 for the discrete map.
The warpage can be accounted for by our choice of initial conditions near
the maximum of expansion, where the approximations used to derive the map
break down. However, the important fine scale structure is laid
down much nearer the singularity, where the map works well,
and this accounts for the agreement in the detailed structure
seen in Figs.~1 and 3. Indeed, a calculation\cite{us} of the information
dimension of Fig.~3 using the uncertainty exponent method
yields $D_{1}=1.86\pm 0.01$, in agreement with the map.

\
\begin{figure}[h]
\vspace{56mm}
\includegraphics{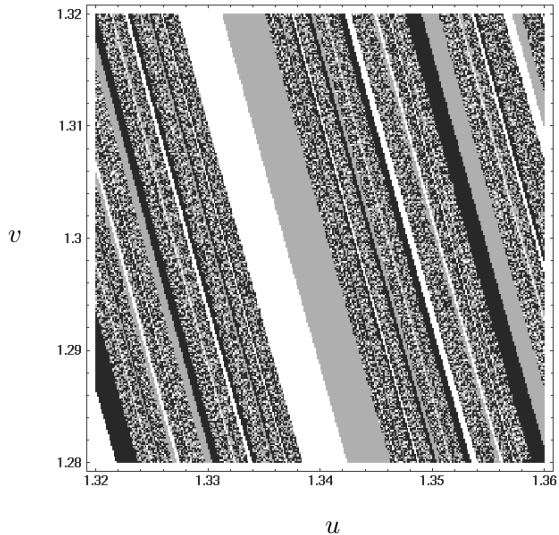}
\vspace{15mm}
\caption{A portion of Fig. 2 magnified 25 times.}
\end{figure} 

\begin{picture}(0,0)
\put(5,150){$v$}
\put(125,40){$u$}
\end{picture}

\vspace*{-4mm}

By exploiting techniques originally developed to study chaotic scattering,
we gain a new perspective
on the evolution of the mixmaster cosmology.
We found a fractal structure, the strange repellor, describes
the chaos well.  The strange repellor is the collection
of all universes periodic in $(u,v)$.
A typical, aperiodic universe will experience a transient age of
chaos if it brushes against the repellor.
The fractal was exposed in both the exact Einstein equations
and the discrete map used to
approximate the evolution. Most
importantly, our fractal approach is independent
of the time coordinate used. An outcome is an outcome no matter how
quickly you get there. Thus, the chaos reflected
in the fractal weave of mixmaster universes is unambiguous.

It would be interesting to extend our study to include inhomogeneous
collapse, and verify the connection between temporal chaos and
spatial turbulence\cite{montani}.
As a final comment, we note that the chaos seen in the mixmaster
system occurs at large curvatures. As is well known,
most of the oscillations and
bounces happen after Planck scale curvatures
have been reached, so quantum effects cannot be ignored.
Nonetheless, our results are consistent with the contention that generic
classical singularities are chaotic. This in turn suggests that
quantum gravity may have to confront quantum chaos.

We thank J. Barrow, C. Dettmann, N. Frankel and P. Ferreira 
for helpful comments. We are grateful to C. Dettmann for letting
us adapt his computer codes.

\end{document}